\documentclass[aps,prl,twocolumn]{revtex4}
\usepackage{amssymb}
\usepackage{amsmath}
\usepackage{amsfonts}

\input{tcilatex}

\begin{document}

\title[Re-entrant superconductivity]{Re-entrant superconductivity in 
Nb/Cu$_{1-x}$Ni$_{x}$ bilayers}
\author{V.~Zdravkov$^{1}$, A.~Sidorenko$^{1}$, G.~Obermeier$^{2}$, 
S.~Gsell$^{2}$, M.~Schreck$^{2}$, C.~M\"{u}ller$^{2}$, S.~Horn$^{2}$, 
R.~Tidecks$^{2}$, L.R.~Tagirov$^{3}$}
\affiliation{$^{1}$Institute of Applied Physics, LISES ASM, Kishinev 2028, 
Moldova\\
$^{2}$Institut f\"{u}r Physik, Universit\"{a}t Augsburg, D-86159 Augsburg,
Germany\\
$^{3}$Solid State Physics Department, Kazan State University, 420008 Kazan,
Russia}

\begin{abstract}
We report on the first observation of a pronounced re-entrant
superconductivity phenomenon in superconductor/ferromagnetic layered
systems. The results were obtained using a
superconductor/ferromagnetic-alloy bilayer of Nb/Cu$_{1-x}$Ni$_{x}$. The
superconducting transition temperature $T_{c}$ drops sharply with increasing
thickness $d_{CuNi}$ of the ferromagnetic layer, until complete suppression
of superconductivity is observed at $d_{CuNi}\approx $4{\ }nm. Increasing
the Cu$_{1-x}$Ni$_{x}$ layer thickness further, superconductivity reappears
at $d_{CuNi}{\approx }$13{\ }nm. Our experiments give evidence for the
pairing function oscillations associated with a realization of the quasi-one
dimensional Fulde-Ferrell-Larkin-Ovchinnikov (FFLO) like state in the
ferromagnetic layer.
\end{abstract}

\volumeyear{year}
\volumenumber{number}
\issuenumber{number}
\eid{identifier}
\date[Date text]{date}
\received[Received text]{date}
\revised[Revised text]{date}
\accepted[Accepted text]{date}
\published[Published text]{date}
\startpage{1}
\endpage{15}
\maketitle



The coexistence of superconductivity (S) and ferromagnetism (F) in a
homogeneous material, described by Fulde-Ferrell and Larkin-Ovchinnikov
(FFLO) \cite{FF,LO}, is restricted to an extremely narrow range of
parameters \cite{Fulde}. So far no indisputable experimental evidence for
the FFLO state exists.

In general, superconductivity and ferromagnetism do not coexist, since
superconductivity requires the conduction electrons to form Cooper pairs
with antiparallel spins, whereas ferromagnetism forces the electrons to
align their spins parallel. This antagonism can be overcome if
superconducting and ferromagnetic regions are spatially separated, as for
example, in artificially layered superconductor/ferromagnet (S/F)
nanostructures (see, e.g. \cite{Chien}, for an early review). The two
long-range ordered states influence each other via the penetration of
electrons through their common interface. Superconductivity in such a
proximity system can survive, even if the exchange splitting energy $E_{ex}$
$\sim k_{B}\theta _{Curie}$ in the ferromagnetic 
layer is orders of magnitude
larger than the superconducting order parameter $\Delta \sim k_{B}T_{c}$,
with $T_{c}$ the superconducting transition temperature. Cooper pairs
entering from the superconducting into the ferromagnetic region experience
conditions drastically different from those in a non-magnetic metal. This is
due to the fact that spin-up and spin-down partners in a Cooper pair occupy
different exchange-split spin-subbands of the conduction band in the
ferromagnet. Thus, the spin-up and spin-down wave-vectors of electrons in a
pair, which have opposite directions, cannot longer be of equal magnitude
and the Cooper pair acquires a finite pairing momentum \cite{Demler}. This
results in a pairing function that does not simply decay as in a
non-magnetic metal, but in addition oscillates on a characteristic length
scale. This length scale is the magnetic coherence length $\xi _{F}$, which
will be specified below.

Various unusual phenomena follow from the oscillation of the pairing wave
function in ferromagnets (see, e.g. the recent reviews \cite%
{G-K-I,L-P,Buzdin05} and references therein). A prominent example is the
oscillatory S/F proximity effect. It can be qualitatively described using
the analogy with the interference of reflected light in a Fabry-P\'{e}rot
interferometer at normal incidence. As the conditions change periodically
between constructive and destructive interference upon changing the
thickness of the interferometer, the flux of light through the interface of
incidence is modulated. In a layered S/F system the pairing function flux is
periodically modulated as a function of the ferromagnetic layer thickness 
$d_{F}$ due to the interference. As a result, the coupling between the S and
F layers is modulated, and $T_{c}$ oscillates as a function of $d_{F}$.

The most spectacular evidence for the oscillatory proximity effect would be
the detection of the re-entrant behavior of the superconducting transition
temperature as a function of $d_{F}$, which has been predicted theoretically 
\cite{Khus,Tag1,Tag2}. There is a sole report on the superconductivity
re-entrance as a function of the ferromagnetic layer thickness in Fe/V/Fe
trilayers \cite{Garif02}. Due to the very small thickness of the iron
layers, at which the re-entrance phenomenon is expected (0.7-1.0 {nm}, i.e.
2-4 monolayers of iron only), the number of the experimental points 
$T_{c}(d_{F})$ is very small, with a large scattering of the results.

The oscillation length $\xi _{F}=\hslash v_{F}/E_{ex}$ in strong
ferromagnets, like iron, nickel or cobalt, is extremely short, because the
exchange splitting energy $E_{ex}$ of the conduction band is in the range
0.1-1.0 eV \cite{Chien}. Here, $v_{F}$ is the Fermi velocity in the F
material and $\hslash $ Planck's constant. Ferromagnetic alloys, with 
$E_{ex} $ an order of magnitude smaller, allow the observation of the effect
at larger thicknesses $d_{F}$ of about 5-10 nm. Such layers can be easier
controlled and characterized. Another advantage using ferromagnetic alloys
is that for a long-wavelength oscillation the atomic-scale interface
roughness has no longer a decisive influence on the extinction of the $T_{c}$
oscillations.

The S/F proximity effect has not only been studied using elemental
ferromagnetic materials, but also for various ferromagnetic alloys 
\cite{Attanasio1,Lohney1,Bader1,Ryazanov2,Potenza,Kim,Attanasio2}. A
non-monotonic dependence of $T_{c}$ \textit{vs}. $d_{F}$ has been observed.
In the present work, Nb was chosen as a superconductor and a 
Cu$_{1-x}$Ni$_{x}$ alloy with $x\approx 0.59$ for the ferromagnetic layer.  
In this alloy the magnetic momentum and Curie temperature show an 
almost linear dependence on the Ni concentration \cite{Vonsovsky}. 
For $x\approx 0.59$ we find $\theta _{Curie}\approx 170$ K.

The samples were prepared by magnetron sputtering on commercial (111)
silicon substrates at 300 K. The base pressure in the \textquotedblleft
Leybold Z400\textquotedblright\ vacuum system was about 2$\times $10 $^{-6}$
mbar, pure argon (99.999\%, \textquotedblleft Messer
Griesheim\textquotedblright ) at a pressures of $8\times 10^{-3}$ mbar was
used as sputter gas. Three targets, Si, Nb and Cu$_{1-x}$Ni$_{x}$ (75 mm in
diameter), were pre-sputtered for 10-15 minutes to remove contaminations
from the targets as well as to reduce the residual gas pressure from the
chamber during the pre-sputtering of Nb, which acts as a getter material.
Then, we first deposited a silicon buffer layer, using a RF magnetron to
generate a clean interface for the subsequently deposited niobium layer. To
average over spatial differences of the sputtering characteristics, we moved
the target during the DC sputtering process of the Nb layer, obtaining a
smooth Nb film of constant thickness $d_{Nb}$. The average growth rate of
the Nb film was about 1.3 nm/sec, while the rate of the sputtering process
was adjusted to 4 nm/sec, to reduce the amount of contaminations gettered
into the Nb film. The Cu$_{1-x}$Ni$_{x}$ target \cite{Ryazanov3} was RF
sputtered (rate 3 nm/sec) resulting in the same composition of the alloy in
the film. As in our previous work \cite{Sidorenko} we deposited a
wedge-shaped ferromagnetic layer to obtain a series of samples with varying
ferromagnetic Cu$_{1-x}$Ni$_{x}$ layer thickness. To prepare this wedge, the
80 mm long and 7 mm wide silicon substrate was mounted at a distance of 4.5
cm from the Cu$_{1-x}$Ni$_{x}$ target symmetry axis to utilize the intrinsic
spatial gradient of the deposition rate. To prevent the degradation of the
Nb/Cu$_{1-x}$Ni$_{x}$ bilayers at atmospheric conditions, the bilayers were
coated by a silicon layer of about 5 nm thickness. A sketch of the resulting
wedge-like samples is presented in the inset of Fig. 1. Samples of equal
length and width were cut from the wedge to obtain a set of 2.5 mm wide
strips with varying Cu$_{1-x}$Ni$_{x}$ layer thickness. Aluminum wires of 50 
$\mu $m in diameter were then attached to the strips by ultrasonic bonding
for four-probe resistance measurements. Two batches of samples were
prepared, one with $d_{Nb}\approx 7.3$ nm (S15), the other with $%
d_{Nb}\approx 8.3$ nm (S16).

After characterizing the samples by resistance measurements, Rutherford
backscattering spectrometry (RBS) has been used to evaluate the thickness of
the Nb and Cu$_{1-x}$Ni$_{x}$ layers as well as to check the concentration
of Cu and Ni in the deposited alloy layers (Fig. 1). The applicability of
this method for thickness determination has been demonstrated in our
previous work \cite{Sidorenko}. It allows determining the thickness (via the
areal density) of the layers with an accuracy of $\pm 3\%$ for Cu$_{1-x}$Ni 
$_{x}$ on the thick side of the Cu$_{1-x}$Ni$_{x}$ wedge, and $\pm 5\%$ for
Nb and Cu$_{1-x}$Ni$_{x}$ on the thin side of the wedge. The measurements
were performed with 3.5 MeV He$^{++}$ ions delivered by a tandem
accelerator. The backscattered ions were detected under an angle of 170$%
^{\circ }$ with respect to the incident beam by a semiconductor detector. In
order to avoid channeling effects in the Si substrate, the samples were
tilted by 7$^{\circ }$ and azimuthally rotated during the measurement. The
spectra were simulated using the RUMP computer program \cite{Doolittle}.
From the deduced elemental areal densities of Nb and Cu$_{1-x}$Ni$_{x}$
alloy the thickness of the two layers was calculated using the densities of
the respective metals. The results for the layer thickness and Cu$_{1-x}$Ni 
$_{x}$ alloy composition as a function of position on the substrate of batch
S15 are shown in Fig. 1. The Ni concentration in the Cu$_{1-x}$Ni$_{x}$
layer is nearly constant showing a slight increase towards the thick side of
the wedge. The thickness of the Nb layer is nearly constant along the wedge, 
$d_{Nb}$(S15)$\approx 7.3$ nm.

The resistance measurements were performed in a $^{3}$He cryostat and a
dilution refrigerator. The standard DC four-probe method was used, applying
a sensing current of 10 ${\mu }$A in the temperature range 0.4 K-10 K and of
2 $\mu $A in the range 10 mK-1.0 K, respectively. The polarity of the
current was alternated during the resistance measurements to eliminate
possible thermoelectric voltages. The superconducting critical temperature 
$T_{c}$ was determined from the midpoints of the $R(T)$ curves at the
superconducting transition (Fig. 2). The transition width (defined by the
temperature interval in which the resistance changes from 0.1$R_{n}$ to 
0.9$R_{n}$, with $R_{n}$ the normal state resistance just above the 
transition) was below 0.2 K for most of the investigated samples. 
The shift between transition measured for increasing and decreasing 
temperature was smaller than 15 mK.

Figure 3 demonstrates for two values of the Nb layer thickness 
($d_{Nb}\approx 8.3$ nm in Fig. 3a and $d_{Nb}\approx 7.3$ nm 
in Fig. 3b) the dependence of the superconducting 
transition temperature on the thickness of
the Cu$_{1-x}$Ni$_{x}$ layer. For specimens with $d_{Nb}\approx 8.3$ nm the
transition temperature $T_{c}$ reveals a non-monotonic behavior with a deep
minimum at about $d_{CuNi}\approx 7.0$ nm. For $d_{Nb}\approx 7.3$ the
transition temperature decreases sharply upon increasing the ferromagnetic 
Cu$_{1-x}$Ni$_{x}$ layer thickness, till $d_{CuNi}\approx 3.8$ nm. Then, in
the range $d_{CuNi}\approx 4.0-12.5$ nm, the superconducting transition
temperature vanishes (at least $T_{c}$ is lower than the lowest temperature
measured in our cryogenic setup, \textit{i.e.} 40 mK). For $d_{CuNi}>12.5$
nm a superconducting transition is observed again with $T_{c}$ increasing up
to about 2 K. This phenomenon of re-entrant superconductivity is the most
important finding of the present study.

For the regions of values of $d_{CuNi}$ for which $T_{c}$ changes rapidly,
the transition width is broader than 0.2 K, and the $R(T)$ curve appears
slightly asymmetric with respect to the midpoint of the transition. This is
probably due to the small variation of the thickness of the ferromagnetic
layer within each sample, since they were cut from a wedge as described
above.

To compare the prediction of the theory for the $T_{c}(d_{F})$ dependence of
Tagirov \cite{Tag1} with our experimental results, we followed closely the
fitting procedure described in detail in reference \cite{Sidorenko}. The
calculated curves of $T_{c}(d_{F})$ in Fig. 3 (a) and (b) agree
qualitatively well with the measured values. The electron mean free path 
$l_{F}\approx 15$ nm for the ferromagnetic material used for 
the calculations appears surprisingly long for an alloy, 
in particular, since a value of $l_{F}\approx 4.4$ nm, was 
inferred from resistivity measurements on a Cu$_{1-x}$Ni$_{x}$ 
alloy with $x\approx 0.51$ \cite{Potenza}. The reason could
be a more complicated character of the diffusion of Cooper pairs in the F
material and its temperature dependence in such type of S/F-system than
considered by the present version of the theory. The small cusp-like
structure in the $T_{c}(d_{CuNi})$ dependence at $d_{CuNi}\approx 1$ nm
cannot be explained by the existing theoretical approaches, \textit{i.e.}
neither the single-mode nor the multi-mode approximation \cite{Fominov,Chun}
can account for this structure. We presume that the case $d_{CuNi}<\xi
_{F},l_{F}$ needs special consideration in the framework of the
\textquotedblleft pure limit\textquotedblright\ theoretical approach 
\cite{Tag2}.

In conclusion, we present the first conclusive experimental observation of
re-entrant behavior of superconductivity and large-amplitude oscillations of
the superconducting $T_{c}$, in two series of superconductor/ferromagnet
bilayers with constant Nb layer thickness ($d_{Nb}\approx 7.3$ nm and 
$d_{Nb}\approx 8.3$ nm) as a function of the thickness of a 
Cu$_{1-x}$Ni$_{x}$ ($x\approx 0.59$) alloy layer.

The authors are grateful to V.~Ryazanov and V.~Oboznov for stimulating
discussions and cooperation, and to Yu. Shalaev for technical assistance in
constructing the target-holder movement setup. The work was partially
supported by INTAS (grant YSF 03-55-1856) and BMBF (project MDA02/002).

\textbf{Figure captions}

Fig. 1. The results of a Rutherford backscattering spectrometry (RBS)
investigation: S15 batch, $d_{Nb}\approx 7.3$ nm.

Fig. 2. Typical resistive transitions of the investigated samples. Solid
lines are a guide to the eye.

Fig. 3. Nonmonotonous $T_{c}$ ($d_{F}$) dependence for Nb/Cu$_{x}$Ni$_{1-x}$
bilayers: (a) d$_{Nb}\approx 8.3$ nm; (b) d$_{Nb}\approx 7.3$ nm. Transition
widths are within the point size if error bars are not visible. Solid lines
are theoretical curves (see text).

\end{document}